\documentstyle[eqsecnum,aps,prd]{revtex}
\begin{document}
\thispagestyle{empty}
\thispagestyle{empty}
{\baselineskip0pt
\leftline{\large\baselineskip16pt\sl\vbox to0pt{\hbox{DAMTP} 
               \hbox{\bf\it University of Cambridge}\vss}}
\rightline{\large\baselineskip16pt\rm\vbox to20pt{
               \hbox{DAMTP-1999-77}
               \hbox{UTAP-330}
               \hbox{RESCEU-99-19}           
               \hbox{\today}
\vss}}%
}
\vskip15mm
\begin{center}
{\large\bf Dilatonic Probe, Force Balance and Gyromagnetic Ratio}
\end{center}

\begin{center}
{\large Tetsuya Shiromizu 
\footnote{JSPS Postdoctal Fellowship for Research Abroad}} \\
\vskip 3mm
\sl{DAMTP, University of Cambridge \\ 
Silver Street, Cambridge CB3 9EW, UK \\
\vskip 5mm
Department of Physics, The University of Tokyo, Tokyo 113-0033, 
Japan \\
and \\
Research Centre for the Early Universe(RESCEU), \\ 
The University of Tokyo, Tokyo 113-0033, Japan
}
\end{center}

\begin{center}
{\it to be published in Phys. Lett. {\bf B}}
\end{center}
\begin{abstract} 
\baselineskip25pt
Following the Papapetrou-Dixon-Wald 
procedure we derive the equation of motion for 
a dilatonic test body(probe) with the dilaton coupling 
$\alpha = {\sqrt {p/(2+p)}}$ in four dimension. 
Since the dilatonic freedom sometimes comes from extra dimensions, it is 
best to derive the EOM by a dimensional reduction from $(p+4)-$dimensions. 
We discuss about the force 
balance up to the gravitational spin-spin interactions via the probe
technique. The force balance condition yields the saturation of 
a Bogomol'nyi bound and the gyromagnetic ratio of the test body. 
\end{abstract}
\vskip1cm


\baselineskip25pt
\section{Introduction}

BPS solitons are an important object and has 
comprehensive features from the view points of 
superstring theory and black hole physics. Great interest is its 
force cancellation. By its virtue multi-soliton solution 
can exist and the existence helps us to construct new 
solutions which might be related to a string state. 
The simplest example is the Majumdar-Papapetrou(MP)  
solution\cite{MP} in general relativity or $N=2$ supergravity\cite{Tod}.  
What we should do in order to check the 
force balance is to investigate the motion of a charged test body on an 
isolated soliton, for example, an extremal Reissner-Nordstr\"om
solution. 
We should observe that the electrostatic 
repulsive force is balanced by the gravitational attractive one 
when $q=m$. 

It is, however, natural to consider rotating solitons in general case. 
It is well known that the Israel-Wilson-Perjes(IWP) solution\cite{PIW} is
stationary one which becomes the MP solution at the static limit.  
The balance between the gravitational spin-spin interaction and
magnetic dipole-dipole force was confirmed by using a spinning 
test body\cite{KT}. The gyromagnetic ratio turns out to be $g=2$ 
which is same with the Kerr solution. Since the body has the spin it does
not follow geodesic motion. Unfortunately, the ISW solution is not extreme
and has naked  singularities except in the static limit\cite{Hartle}.  

In the heterotic string theory on a torus, similar $D-$dimensional 
stationary charged solution has been found 
through $O(d-1,d-1)$ transformation from an uncharged Kerr solution
\cite{HS}\cite{Sen}. For $D > 5$  the 
extreme limit saturates a Bogomol'nyi bound without naked singularities 
and then we can 
obtain a regular multi-soliton solution\cite{HS}. 
Here we remind you that the extreme limit of 
non-dilatonic Kerr solution does not exists for even $D>5$. But now the
solution contains the dilaton sector and the situation is changed. 
We also be able to obtain a four dimensional 
solution by taking doubly periodic arrays of extremal black holes 
in six dimensions. Thus it is worth to study the force balance in dilatonic 
or stringy multi-black hole systems. 

In this paper, as a first step, we derive the equation of motion for 
a dilatonic test body with the  
dilaton coupling $\alpha={\sqrt {p/(2+p)}}$ in four dimension. 
This theory can be easily obtained from  
the dimensional reduction of ($p+4$)-dimensional 
Maxwell-Einstein theory. The motion and force balance of a  
test body without any moments, i.e. {\it test particle}, has been 
investigated in ref. \cite{Maki}. 

The rest of this paper is organised as follows. In Sec. II we briefly
review the Papapetrou-Dixon-Wald procedure\cite{Papa}\cite{WD}\cite{Wald} 
in higher dimensions and give the equation of 
motion for a test body. In Sec. III, we rewrite the above formal
equation in terms of four dimensions and discuss about the force balance up 
to the gravitational spin-spin interaction and give the force balance 
condition. It is well known that the force balance between the
gravitational, dilatonic and electrostatic
forces yields the saturation of the Bogomol'nyi bound. 
Moreover the condition supplies the gyromagnetic ratio of the test
body balancing with the central soliton. Finally we give summary and 
discussion in Sec. IV. 

\section{Equation of Motion}

The ($p+4$)-dimensional Einstein-Maxwell theory gives
the Einstein-Maxwell-Dilaton theory 
with the coupling $\alpha={\sqrt {p/(p+2)}}$ in 
four dimension. We follow the argument given in ref. \cite{GHT}. 
Let the metric 
$G_{MN}$ of  ($p+4$)-dimensional space-time to be 
%
\begin{eqnarray}
ds^2_{4+p}=G_{MN}dx^M dx^N=e^{2\alpha\phi}g_{\mu\nu}dx^\mu dx^\nu+
e^{2(\alpha -\alpha^{-1})\phi}(d{\chi}_1^2+\cdots +d{\chi}_p^2),
\end{eqnarray}
%
where $\alpha={\sqrt {p/(p+2)}}$ and $\phi$ is a scalar function of
$\lbrace x^\mu \rbrace$. The Geek indices runs from $0$ to $3$. 
As you can easily see, the Einstein-Hilbert 
action in ($p+4$)-dimension yields 
%
\begin{eqnarray}
S=\frac{1}{16\pi G}\int d^4x {\sqrt {-g}}[R_g-2(\nabla
\phi)^2-e^{-2\alpha \phi}F^2] \label{eq:action}
\end{eqnarray}
%
in terms of four dimension. Here $R_g$ is the Ricci scalar of 
the metric $g_{\mu\nu}$. 

Following the Papapetrou-Dixon-Wald
procedure\cite{Papa}\cite{WD}\cite{Wald},  
we can instantly written down the
equation of motion from the conservation low of the energy-momentum 
tensor in higher dimension, $\nabla^MT_{MN}=0$. The result is as follows,
%
\begin{eqnarray}
v^M \nabla_M p^N =\frac{1}{2}{}^{(4+p)}R^N_{M I J}v^M S^{IJ}
+qv^MF^N_M+\frac{gq}{4m}S^{IJ} \nabla^N F_{IJ} \label{eq:main}
\end{eqnarray}
%
%
\begin{eqnarray}
v^M \nabla_M S^{IJ}=2 p^{[I} v^{J]} +\frac{gq}{2m}S^{M [J}F^{I]}_M, 
\end{eqnarray}
%
where $g$ is the gyromagnetic ratio of the spinning test body. $p^M$
and $v^M$ are $M-$momentum and $M-$velocity of a test body which has
mass $m$, charge $q$ and angular momentum tensor $S^{MN}$. 
In the above we imposed the ``supplementary condition'' on the 
test body,  
%
\begin{eqnarray}
p^M S_{MN}=0
\end{eqnarray}
%
which determines the motion of the test body and specifies  
`the center of mass'\cite{Wald}. 
We find the fact that $p^M$
is not always proportional to $v^M$ because of the existence of 
spin. In fact the relation turns out in the form
%
\begin{eqnarray}
v^M  =  f \Bigl[  p^M-\frac{1}{p^I p_I }
\frac{\Bigl[\frac{1}{2}R_{NIJK}S^{JK}-q F_{NI}\Bigr]p^I S^{MN} 
}{1-\frac{1}{p^J p_J}[\frac{1}{4}R_{KLPQ}S^{KL}
S^{PQ}-\frac{1}{2}qS^{KL}F_{KL}]} \Bigr],\label{eq:vp}
\end{eqnarray}
%
where $f=(v^M p^M)/(p^N p_N)$. However, $p^M \propto v^M $ if one 
ignores the higher order terms at the spacelike asymptote. 
%
%
%
From eq. (\ref{eq:main}) we obtain 
%
\begin{eqnarray}
v^M \nabla_M Q_\xi = -\frac{gq}{4m}S^{MN} \pounds_\xi F_{MN}
\end{eqnarray}
%
where
%
\begin{eqnarray}
Q_\xi=(p^M +q A^M)\xi_M +\frac{1}{2}S^{MN}\nabla_N \xi_M.
\end{eqnarray}
%
Hence $Q_\xi$ is conserved if $\pounds_\xi F_{MN}=0$ holds. The
condition $\pounds_\xi F_{MN}=0$ holds on the stationary exact solution. 

As a next step we will rewrite the above equation in terms of 
four dimension. We will do that for examples given in the next section. 

\section{Force Balance}

The force balance between solitons is important features because 
it guarantees the existence of multi-soliton and this gives a 
new black hole solution by taking some periodic arrays of solitons 
which might be related to a string state. 

First of all we derive the equation of motion for a test body without 
spin on a static charged dilatonic solution\cite{dilaton1}. 

\subsection{Non-Rotating Case}

In this case we can take $S^{IJ}=0$ and 
the equation of motion is simply given by the geodesic motion in 
($p+4$)-dimension; 
%
\begin{eqnarray}
\frac{d}{d\tau} p^M=-\Gamma^N_{MI}v^Mp^I+qF^N_Iv^I.
\end{eqnarray}
%
Here we are interest in the motion in four dimensional space-time 
and it is natural to assume that the extra dimensions components of the 
velocity $v^M$ vanish. 
From eq. (\ref{eq:vp}) we see that $p^M$ is proportional to $v^M$;
%
\begin{eqnarray}
p=p^I\partial_I=p^\mu \partial_\mu=f^{-1} v,
\end{eqnarray}
%
where $v^\mu=dz^\mu(\tau)/d\tau$ and $g(v,v)=-1$. From the above 
equation of motion 
%
\begin{eqnarray}
v^M \nabla_M(p^Ip_I)=-\partial_\tau(f^{-2} e^{2\alpha \phi})=0
\end{eqnarray}
%
holds. Then $f \propto e^{\alpha \phi}$ and 
%
\begin{eqnarray}
p^I=me^{-\alpha \phi}v^I.
\end{eqnarray}
%
After arranging of several terms we obtain the final form as 
%
\begin{eqnarray}
v^\mu \nabla_\mu (e^{\alpha \phi}v^\nu)+\nabla^\nu e^{\alpha \phi}
=\frac{q}{m}F^\nu_\mu v^\mu. 
\end{eqnarray}
%
This equation is also derived from the variational principle of the
action given by \cite{Maki};
%
\begin{eqnarray}
S_4=\int d \tau \Bigl[ m{\sqrt {-e^{2\alpha
\phi}g_{\mu\nu}\frac{dx^\mu}{d\tau}
\frac{dx^\nu}{d\tau} }}+eA_\mu \frac{dx^\mu}{d\tau}   \Bigr]. 
\end{eqnarray}
%
We have the constant of motion like energy per mass, ${\cal E}$;
%
\begin{eqnarray}
{\cal E}:=-\Bigl(e^{\alpha \phi}v^\mu+\frac{q}{m}A^\mu \Bigr)\xi_\mu 
\label{eq:energy}
\end{eqnarray}
%
and
%
\begin{eqnarray}
\pounds_v {\cal E}=0
\end{eqnarray}
%

The metric, vector potential and dilation field of the static black hole 
solution for the action (\ref{eq:action}) are given by\cite{dilaton1}  
%
\begin{eqnarray}
ds^2=g_{\mu\nu}dx^\mu dx^\nu = -\frac{\Delta}{\sigma^2}dt^2
+\sigma^2\Bigl[\frac{1}{\Delta}dr^2+r^2 d\Omega_2^2 \Bigr]\label{eq:metric1}
\end{eqnarray}
%
%
\begin{eqnarray}
A=-\frac{Q}{r}dt~~~~~{\rm and}~~~~~~e^{2\alpha \phi}=\sigma^2,
\end{eqnarray}
%
where
%
\begin{eqnarray}
\Delta=\Bigl(1-\frac{r_-}{r}\Bigr)\Bigl(1-\frac{r_+}{r} \Bigr)
~~~~~{\rm and}~~~~~
\sigma^2=\Bigl(1-\frac{r_-}{r} \Bigr)^{2\alpha^2/(1+\alpha^2)}.
\end{eqnarray}
%
$r_{\pm}$ are related to the ADM mass and total charge as follows, 
%
\begin{eqnarray}
2M=r_++\frac{1-\alpha^2}{1+\alpha^2}r_-~~~~~{\rm and}~~~~~
Q^2=\frac{r_+r_-}{1+\alpha^2}.
\end{eqnarray}
%

Inserted metric (\ref{eq:metric1}) into eq. (\ref{eq:energy}) we
obtain
%
\begin{eqnarray}
{\cal E}=\frac{\Delta}{\sigma^2}e^{\alpha \phi} \frac{dt}{d\tau}
+\frac{q}{m}\frac{Q}{r}
\end{eqnarray}
%
and then
%
\begin{eqnarray}
\Bigl( \frac{dt}{d\tau} \Bigr)^2= \frac{\sigma^2}{\Delta^2} \Bigl( 
{\cal E}-\frac{qQ}{mr}\Bigr)^2. \label{eq:dtdtau}
\end{eqnarray}
%
From $g(v,v)=-1$ and eq. (\ref{eq:dtdtau}) 
%
\begin{eqnarray}
\Bigl(\frac{dr}{d\tau}\Bigr)^2+V(r)=0,
\end{eqnarray}
%
where
%
\begin{eqnarray}
V(r)=\frac{1}{\sigma^2}\Bigl[-\frac{r_++r_--2qQ/m}{r} 
+\frac{r_+r_--(qQ/m)^2}{r^2} \Bigr]
\end{eqnarray}
%

When $r_+=r_-$, that is, 
%
\begin{eqnarray}
Q=(1+\alpha^2)^{1/2}M~~~~~{\rm and}~~~~~q=(1+\alpha^2)^{1/2}m,
\end{eqnarray}
%
the potential exactly vanishes, that is, $V(r)=0$ for arbitrary 
$r$. In the dilatonic gravity 
theory, the Bogomol'nyi type bound($M \geq (1+\alpha^2)^{-1/2}|Q|$) has been 
proven\cite{BB}. So we realise again that 
the extreme limit saturates  the Bogomol'nyi
bound. We also be able to check the 
force balance from the equation of motion directly. For simplicity,
hereafter, we take the 
extreme limit. The metric is written as 
%
\begin{eqnarray}
ds^2=g_{\mu \nu}dx^\mu dx^\nu=-V(R)^{-1}dt^2+V(R)d{\bf X}^2,
\end{eqnarray}
%
where $V(R)=(1+r_+/R)^{2/(1+\alpha^2)}=(1+r_+/R)^{(p+2)/(p+1)}$ and 
$R=r-r_+={\sqrt{\delta_{ij}X^iX^j}}$. Then
%
\begin{eqnarray}
\frac{d^2x^i}{d \tau^2}=-\Gamma^i_{00}-\alpha \partial_i \phi
+\frac{q}{m}
e^{-\alpha \phi}\partial_i A_0+O(\frac{1}{R^3})=O(\frac{1}{R^3})
\end{eqnarray}
%
because of $\Gamma^i_{00} \simeq \frac{r_+}{1+\alpha^2}\frac{X^i}{R^3}$, 
$\partial_i \phi \simeq \frac{\alpha r_+}{1+\alpha^2}\frac{X^i}{R^3}$ and 
$\frac{q}{m}\partial_i A_0 \simeq r_+ \frac{X^i}{R^3}$. This tells us 
the cancellation between the gravitational, dilatonic 
and electrostatic forces. 

We remember that the above extreme dilatonic solution with odd $p$ 
can be derived 
from dense arrays of infinite multi-black holes solution with the same mass along 
the $\chi$-direction in the ($p+4$)-dimensional Einstein-Maxwell 
theory\cite{compact}. 
The metric of multi-black holes solution is given by  
%
\begin{eqnarray}
ds^2_{4+p}=-U^{-2}dt^2+U^{2/(p+1)}d{\bf x}^2 +U^{2/(p+1)}
(d\chi_1^2+ \cdots +d\chi_p^2) \label{eq:pblane},
\end{eqnarray}
%
where $U=1+\sum_{i=1}^p m_i/(|{\vec x}-{\vec x}_i|^{p+1} 
+|{\vec \chi}-{\vec \chi}_i |^{p+1})$. The dense arrays 
along $p$'s $\chi$ direction leads us
%
\begin{eqnarray}
U=1+\sum_{i=1}^{p}\frac{\mu}{|\vec{x}|^{p+1}+|\vec{\chi}-\vec{\chi}_i|^{p+1}}
=1+\frac{M}{r}. 
\end{eqnarray}
%
Note that the metric (\ref{eq:pblane}) can be written as 
%
\begin{eqnarray}
ds^2_{4+p}& = & U^{-p/(1+p)}\Bigl[-U^{-(p+2)/(p+1)}dt^2+U^{(p+2)/(p+1)}
d{\bf x}^2 \Bigr]+U^{2/(p+1)}(d\chi_1^2+ \cdots + d\chi_p^2) 
\nonumber \\
& = & U^{-p/(1+p)} g_{\mu\nu}dx^\mu dx^\nu
+U^{2/(p+1)}(d\chi_1^2+ \cdots + d\chi_p^2) .
\end{eqnarray}
%
This is black $p$-brane. Thus the motion of test body is equivalent
with the geodesic motion for perpendicular direction to the surface 
where arrays(black $p$-brane) are.

\subsection{Rotating Case}

For purpose of this paper it is enough to pick up the first order
of the angular momentum (slow rotating approximation). 
The general solution is given by\cite{dilaton2} 
%
\begin{eqnarray}
ds^2=-\frac{\Delta}{\sigma^2}+\sigma^2 \Bigl[ \frac{1}{\Delta}dr^2+r^2
d \Omega_2^2 \Bigr]-2af(r){\rm sin}^2 \theta dt d\phi
\end{eqnarray}
%
%
\begin{eqnarray}
f(r) & = & \frac{(1+\alpha^2)^2}{(1-\alpha^2)(1-3 \alpha^2)}
\Bigl(\frac{r}{r_-} \Bigr)^2
\Bigl( 1-\frac{r_-}{r}\Bigr)^{2\alpha^2/(1+\alpha^2)} \nonumber \\
& & -\Bigl(1-\frac{r_-}{r} \Bigr)^{(1-\alpha^2)/(1+\alpha^2)} 
\Bigl[ 1+\frac{(1+\alpha^2)^2}{(1-\alpha^2)(1-3\alpha^2)}
\Bigl(\frac{r}{r_-}\Bigr)^2+\frac{1+\alpha^2}{1-\alpha^2}\frac{r}{r_-}
-\frac{r_+}{r} \Bigr]
\end{eqnarray}
%
and
%
\begin{eqnarray}
A=\frac{Q}{r}dt-a{\rm sin}^2 \theta \frac{Q}{r}d \varphi
\end{eqnarray}
%
As, at first sight, the function 
$f(r)$ seems to have a bad behaviour at $p=1$,
we will consider the case with $p=1$ separately. 

The apparent bad behaviour of the function $f(r)$ does not yield any
troubles and we obtain the expression  
%
\begin{eqnarray}
f(r)=-\Bigl(1-\frac{r_-}{r}
\Bigr)^{1/2}\Bigl[ \Bigl(1-\frac{r_+}{r}+\frac{2r}{r_-} 
\Bigr)-\frac{2r^2}{r_-^2}{\rm ln}\Bigl(1-\frac{r_+}{r} \Bigr) \Bigr]
\end{eqnarray}
%
at the limit $p=1$. Since we have $f(r)\simeq
\frac{5}{3}\frac{r_+}{R}$ at the extreme 
limit($r_+=r_-$), the metric and vector potential become 
%
\begin{eqnarray}
ds^2=-(1+r_+/R)^{-3/2}dt^2+(1+r_+/R)^{3/2}d{\bf X}^2+\frac{20}{3}
\frac{1}{R^3}\epsilon_{ijk}J^jX^kdtdX^i
\end{eqnarray}
%
and
%
\begin{eqnarray}
A \simeq \frac{Q}{R}dt + \frac{2Q}{r_+R^3}\epsilon_{ijk}J^jX^k dX^i,
\end{eqnarray}
%
where $|J|=\frac{ar_+}{2}$. As 
%
\begin{eqnarray}
R^i_{MIJ}v^MS^{IJ} \simeq {}^{(4+p)}R^i_{0jk}S^{ij}
\end{eqnarray}
%
and
%
\begin{eqnarray}
{}^{(p+4)}R^i_{0jk}& = & {}^{(4)}R^i_{0jk}-2\alpha^2g_{0[j}\nabla_{k]}\phi
\nabla^i \phi+2 \alpha g_{0[j}\nabla_{k]}\nabla^i \phi 
-\alpha \delta^i_{[j}\nabla_{k]}\nabla_0\phi \nonumber \\
& = & {}^{(4)}R^i_{0jk} +O(1/R^5) = 2\partial_{[j} \Gamma^i_{k]0}+O(1/R^5),
\end{eqnarray}
%
we can evaluate the first and third terms of the right-hand side in
eq. (\ref{eq:main}) as follows,  
%
\begin{eqnarray}
F_{{\rm spin-spin}}^i=-\frac{1}{2}{}^{(4)}R^i_{0jk}S^{jk}=\frac{5}{3}
\frac{\partial}{\partial X^i} \Bigl[ \frac{-{\bf J} \cdot {\bf S}+
3({\bf J}\cdot {\hat {\bf X}}) ({\bf S}\cdot {\hat {\bf X}}) }{R^3}
\Bigr]+O(1/R^5)
\end{eqnarray}
%
and
%
\begin{eqnarray}
F_{{\rm dipole-dipole}}^i=-\frac{gq}{4m}S^{jk}{}^{(5)}\nabla^i F_{jk}
\simeq -\frac{gqQ}{mr_+}\frac{\partial}{\partial X^i}
\Bigl[  \frac{-{\bf J} \cdot {\bf S}+
3({\bf J}\cdot {\hat {\bf X}}) ({\bf S}\cdot {\hat {\bf X}}) }{R^3}
\Bigr]+O(1/R^5),
\end{eqnarray}
%
where ${\hat {\bf X}}={\bf X}/R$. By summing up those we obtain 
%
\begin{eqnarray}
F_{{\rm spin-spin}}^i+F_{{\rm dipole-dipole}}^i=
\Bigl( \frac{5}{3}-\frac{gqQ}{mr_+}\Bigr) \frac{\partial}{\partial X^i}
\Bigl[ \frac{-{\bf J} \cdot {\bf S}+
3({\bf J}\cdot {\hat {\bf X}}) ({\bf S}\cdot {\hat {\bf X}}) }{R^3}
\Bigr]+O(1/R^5),
\end{eqnarray}
%
As we see in the previous subsection, monopole components are  
balanced by each other when $Q=2M/{\sqrt {3}}$ and $q=2m/{\sqrt
{3}}$. Then the balance between the above forces holds when the 
gyromagnetic ratio is $g=5/3$. 

Next we consider cases with $p \neq 1$. The metric and 
vector potential are approximately given by 
%
\begin{eqnarray}
ds^2=-\frac{\Delta}{\sigma^2}+\sigma^2 \Bigl[ \frac{1}{\Delta}dr^2+r^2
d \Omega_2^2 \Bigr]+\frac{4}{R^3}\epsilon_{ijk}J^jX^kdtdX^i
\end{eqnarray}
%
and
%
\begin{eqnarray}
A\simeq
\frac{Q}{R}dt+\frac{3(1+\alpha^2)}{(3+\alpha^2)}\frac{Q}{r_+R^3}
\epsilon_{ijk}J^jX^kdX^i
\end{eqnarray}
%
where $|J|=\frac{(3+\alpha^2)ar_+}{3(1+\alpha^2)}$. 
As the same way as $p=1$ case, we obtain
%
%
%
\begin{eqnarray}
F_{{\rm spin-spin}}^i+F_{{\rm dipole-dipole}}^i=
\Bigl( 1
-\frac{3(1+\alpha^2)}{2(3+\alpha^2)}\frac{gqQ}{mr_+}\Bigr) 
\frac{\partial}{\partial X^i}
\Bigl[ \frac{-{\bf J} \cdot {\bf S}+
3({\bf J}\cdot {\hat {\bf X}}) ({\bf S}\cdot {\hat {\bf X}}) }{R^3}
\Bigr]+O(1/R^5),
\end{eqnarray}
%
Then the balance between the above forces holds when the 
gyromagnetic ratio is $g=\frac{2(3+\alpha^2)}{3(1+\alpha^2)}$. 
Here we realise that the previous value of the gyromagnetic ratio 
in the case with $p=1$ can be combined to the present expression. 
It is remarkable that the value is different from that of the background 
space-time.  According to ref. \cite{dilaton2}, the background
space-time has $g=\frac{6}{3+\alpha^2}$. That is, the test body is 
not like the background space-time. The discrepancy indicates 
the non-existence of spinning multi-soliton solutions in the 
Einstein-Maxwell-Dilaton theory. If and only if $\alpha=p=0$, 
all gyromagnetic ratios are same as $g=2$ and the exact IWP solutions
exists as we expect. 

\section{Summary and Discussion}

In this paper we derived the equation of motion for a spinning charged dilatonic
probe(test body) by using the Paparetrou-Dixon-Wald 
procedure and discuss about the 
force balance. We found that the force balance holds up to 
the spin-spin interaction when $|Q|={\sqrt {1+\alpha^2}}M$, 
$|q|={\sqrt {1+\alpha^2}}m$ 
and the gyromagnetic ratio $g=\frac{2(3+\alpha^2)}{3(1+\alpha^2)}$. The 
former condition saturates the Bogomol'nyi bound and its fact was well 
known. On the other hand, the latter condition has a discrepancy with
the value of the background space-times except in the $p=0$ case. 
This fact might mean that the Einstein-Maxwell-Dilaton theory does not
have the exact spinning multi-soliton solutions. We are aware of the 
multi-soliton solution in $N=4, D=4$ supergravity\cite{susy}  
or heterotic string theory\cite{HS}\cite{Sen}. 
In those theory the additional ingredient is 
the three form field $H_{\mu \nu \alpha}$ which can be expressed by 
the `axion' field, $a$, so that $H_{\mu\nu\alpha}=e^{4\phi}
\epsilon_{\mu\nu\alpha\beta}
\partial^\beta a$ in four dimension. Hence, the resolution to the 
discrepancy may be 
essential to investigate the force balance in so called 
`Einstein-Maxwell-Dilaton-Axion' theory. The investigation on 
the stringy probe, that is, test body with the moment associated 
to the three form field might be important and will be reported in the
next study\cite{TS}. In this paper our
consideration was concentrated on the bosonic sector. Obviously, 
the deeper investigation of the fermion sector is also great interest 
although we can expect a parallel issue naively. 

\vskip1cm

\centerline{\bf Acknowledgment}
The author is grateful to Gary Gibbons and DAMTP relativity group for their 
hospitality. He also thanks M. Spicci for his careful reading of this
manuscript. This study is supported by JSPS(No. 310).


\end{document}